\documentclass[aps, prd, showpacs, superscriptaddress, nofootinbib]{revtex4}
\usepackage{amssymb,amsmath,amsfonts,amsbsy,graphicx,microtype,rotating}

\usepackage{color}

\begin{document}
\title{A possible interpretation of the Higgs mass by the cosmological attractive relaxion}

\author{Fa Peng Huang}
\email{huangfp@ihep.ac.cn}
\affiliation{Theoretical Physics Division, Institute of High Energy Physics, Chinese Academy of Sciences, P.O.Box 918-4, Beijing 100049, P.R.China}

\author{Yifu Cai}
\affiliation{CAS Key Laboratory for Researches in Galaxies and Cosmology, Department of Astronomy, University of Science and Technology of China, Chinese Academy of Sciences, Hefei, Anhui 230026, China}

\author{Hong Li}
\affiliation{Key Laboratory of Particle Astrophysics, Institute of High Energy Physics, Chinese Academy of Science, P.O.Box 918-3, Beijing 100049, P.R.China
}

\author{Xinmin Zhang}
\affiliation{Theoretical Physics Division, Institute of High Energy Physics, Chinese Academy of Sciences, P.O.Box 918-4, Beijing 100049, P.R.China}

\begin{abstract}
Recently, a novel idea~\cite{Graham:2015cka} has been proposed to relax the electroweak hierarchy problem through the cosmological inflation
and the axion periodic potential.
Here, we further assume that only the attractive inflation is needed to explain the light
mass of the Higgs boson, where we do not need a specified periodic potential of the axion field.
Attractive inflation during the early universe drives the Higgs boson mass from the large value in the early universe to the small value at present,
where the Higgs mass is an evolving parameter of the Universe.
Thus, the small Higgs mass can technically originate from the cosmological evolution rather than dynamical symmetry
or anthropics.
Further, we study the possible
collider signals or constraints at future lepton collier and the possible constraints from the muon anomalous magnetic moment.
A concrete attractive relaxion model is also discussed, which is consistent with the data of Planck 2015.
\end{abstract}



\maketitle

\section{Introduction}

Recently,  a novel type of resolution for the electroweak (EW) hierarchy problem~\footnote{
Many mechanisms have been proposed to solve the EW hierarchy problem,
including the extra dimensions theory, the supersymmetry theory, and the compositeness of the Higgs boson~\cite{Dimopoulos:1981zb, Randall:1999ee, ArkaniHamed:1998rs, Susskind:1978ms, Weinberg:1979bn}.
However, all these new models lead to a technically natural EW scale~\cite{'tHooft:1980xb}, since the current experimental data from
the colliders and indirect experiments have put their model parameters into fine-tuned regions.
Or, we can just ignore the EW hierarchy problem if we believe the anthropic principle.} has been proposed (relaxion mechanism) in Ref.~\cite{Graham:2015cka}, which is
very different from the traditional approaches (either weak scale dynamics~\cite{Dimopoulos:1981zb, Randall:1999ee, ArkaniHamed:1998rs, Susskind:1978ms, Weinberg:1979bn} or anthropics).
In this relaxion mechanism, the relevant fields below the cutoff scale are just the standard model (SM) fields plus the axion field  with unspecified  inflation sector being involved, and the Higgs mass is dependent on the axion field~\cite{Graham:2015cka},  which is motivated
from Abbott's field-dependent idea to solve the cosmological constant problem~\cite{Abbott:1984qf}.
Accordingly, the cosmological evolution of the Higgs mass and the specific axion potential choose the EW
scale, which is smaller than the cutoff of the theory.
The highest cutoff relaxed in Ref.~\cite{Graham:2015cka} is about $10^8$ GeV.
This relaxion mechanism  can technically relax the  EW hierarchy problem
and has become
a theoretical highlight in frontier studies of the hierarchy problem and
exploring new physics beyond the SM~\cite
{Strumia:2015dca,Espinosa:2015eda,Hardy:2015laa,Patil:2015oxa,Antipin:2015jia,Jaeckel:2015txa,
Gupta:2015uea,Batell:2015fma,Matsedonskyi:2015xta,Ellis:2015oda,Marzola:2015dia,Choi:2015fiu,
Kaplan:2015fuy,DiChiara:2015euo,Fonseca:2016eoo,Evans:2016htp,Fowlie:2016jlx}.
Especially, this new mechanism opens a new window to understand some puzzles and key
parameters in particle physics from the aspect of cosmological evolution.

Following this cosmological evolution idea, our toy model here takes advantage of the attractive properties of the ``$\alpha$-attractors''~\cite{Kallosh:2013lkr,Kallosh:2013pby,Kallosh:2013hoa,
Ferrara:2013rsa,Kallosh:2013tua,Kallosh:2013yoa,Kallosh:2014rga,Kallosh:2014laa,Galante:2014ifa,Kallosh:2015lwa,Linde:2015uga,Carrasco:2015rva,Carrasco:2015pla}
to fix the Higgs mass at today's value rather than the increasing potential barriers of the axion potential in ~\cite{Graham:2015cka}.
In recent years,  Linde and Kallosh have proposed a broad class of supergravity inflationary models based on conformal symmetry in the Jordan frame, where a universal attractor behavior exists in the Einstein frame \cite{Kallosh:2013lkr,Kallosh:2013pby,Kallosh:2013hoa,
Ferrara:2013rsa,Kallosh:2013tua,Kallosh:2013yoa,Kallosh:2014rga,Kallosh:2014laa,Galante:2014ifa,Kallosh:2015lwa,Linde:2015uga,Carrasco:2015rva,Carrasco:2015pla}. These classes of  supergravity inflationary models are called ``$\alpha$-attractors'', since their potentials involve a free parameter $\alpha$ \cite{Kallosh:2013yoa,Kallosh:2014rga,Kallosh:2014laa,Galante:2014ifa,Kallosh:2015lwa,Linde:2015uga,Carrasco:2015rva,Carrasco:2015pla}.
These inflation models of ``$\alpha$-attractors'' mainly have two classes: one is the so-called T-models with the potential $f^2(\tanh(\phi/(\sqrt{6\alpha}\rm M_{\rm pl})))$ and the other is the so-called E-models with the potential $f^2(1-\exp(-\sqrt{2/3\alpha}\phi/\rm M_{\rm pl}))$ in the Einstein frame. In the limits of small $\alpha$ and large $e$-folding number $\text{N}_e$, these models have the same predictions \cite{Kallosh:2013yoa,Kallosh:2014rga,Kallosh:2014laa,Galante:2014ifa,Kallosh:2015lwa,Linde:2015uga,Carrasco:2015rva,Carrasco:2015pla} corresponding to the central area of the $n_{\textrm{s}}-r$ plane favored by the data of Planck 2015 \cite{Planck:2015xua}.

Our toy model here only tries to provide a possible cosmological interpretation of the light Higgs mass,
and only the attractive inflation field is needed motivated from the above relaxion mechanism and the attractive inflation.
Here, we do not consider the UV-completed theory for a fully natural theory.
In addition, our models may be tested in  particle physics experiments.

In Section 2,
we first describe the cosmological scenario to explain the light mass of the Higgs boson by attractive inflation.
In Section 3,
we study the possible signals or constraints in future lepton colliders and the constraints from the muon anomalous magnetic moment.
Then, in Section 4, we perform a detailed analysis of cosmological perturbations seeded by the inflation fields in
a concrete model.
Section 5 gives a brief summary.

\section{A cosmological scenario to explain the  Higgs mass by the attractive   relaxion}\label{scenario}
In this section, we show a possible cosmological scenario to explain the light mass of the Higgs boson by  attractive inflation.
In our toy model, the field contents below the cutoff scale  are just the SM fields and the inflaton field.
The relevant potential can be written as
\begin{equation}\label{effp}
V(\phi,h)=\frac{\lambda_{SM}}{4}h^4+\frac{(g^2 \phi^2- \text{M}^2 )}{2}h^2+V_{att},
\end{equation}
where  the field $h$ and $\rm \phi$ represent the Higgs field and the inflaton field, respectively.
The second  term  in Eq.(\ref{effp}) generally can be the form of $(f(\phi)-\text{M}^2)h^2/2$.
In this scenario, the Higgs mass in the early universe is field dependent, namely, $m_h^2=g^2 \phi^2-\rm M^2$.
This is just the starting point of our discussion.
We assume that the initial value of the inflation field $\phi$ starts at $\phi \gg \text{M}/g$, where
M represents the cutoff scale in this toy model.
Thus, the mass of the Higgs boson in the early universe is naturally set to be the order of the cutoff scale M.
Here, $V_{att}$ means the attractive potential, which can drive the cosmological inflation and fix the current
Higgs mass. Interestingly, the potentials in a broad class of  supergravity inflation models (the so-called ``$\alpha$-attractors'')
\cite{Kallosh:2013yoa,Kallosh:2014rga,Kallosh:2014laa,Galante:2014ifa,Kallosh:2015lwa,Linde:2015uga,Carrasco:2015rva,Carrasco:2015pla}
can just satisfy the requirements.
One class of potentials
is given by \cite{Kallosh:2013yoa,Kallosh:2014rga,Kallosh:2014laa,Galante:2014ifa,Kallosh:2015lwa,Linde:2015uga,Carrasco:2015rva,Carrasco:2015pla}
\begin{equation}\label{vt}
V_{att}=V_{\text{T}}=\text{M}^4\tanh^{2\text{T}}\biggl({\phi-\phi_c\over \sqrt{6\alpha}{\rm M_{\rm pl}}}\biggr),
\end{equation}
where $\text{M}_{\rm pl}$ is the reduced Planck mass.
This class of potential is just the potential in the T-models of ``$\alpha$-attractors''.
The potential $V_\text{T}$ can be obtained from canonical Kahler potential such as $(\Phi-\bar{\Phi})^2$ in the
supergravity model.
The  non-minimal coupling case between  the inflaton field and  the gravitational field
can also lead to the same predictions  when compared to this type of attractive potential in some limits.
The other type of attractive potential can be written as~\cite{Kallosh:2013yoa,Kallosh:2014rga,Kallosh:2014laa,Galante:2014ifa,Kallosh:2015lwa,Linde:2015uga,Carrasco:2015rva,Carrasco:2015pla}
\begin{equation}\label{ve}
V_{att}=V_{\text{E}} = \text{M}^{4} \left(1-e^{-{\sqrt {2\over 3 \alpha}}(\phi-\phi_c)/\text{M}_{\rm pl}}\right)^{2\text{E}} \ .
\end{equation}
This potential can be motivated by considering a vector rather than a chiral
multiplet for the inflation models in supergravity, and is just the potential in the so-called E-model.
Here, the power exponents T and E in Eqs.~(\ref{vt}) and (\ref{ve}) are integers and $\alpha$ is the free parameters.
$\phi_c$ is a constant, which is  related to the current Higgs boson mass.

We take the potential $V_\text{T}$ as an example to illuminate the cosmological origin of  the light Higgs boson mass.
Firstly, under the slow-roll approximation, the spectral index $n_s$,
the tensor-to-scalar ratio $r$, and the e-folding number $\text{N}_e$ as functions of $\phi$ can be obtained:
\begin{eqnarray}
&&n_s=
1-{1\over 3\alpha}\biggl[4T~{\rm sech}^2{\sqrt{1\over 6\alpha}{\phi-\phi_c\over \text{M}_{\rm pl}}} \nonumber \\
&&+8\text{T}(1+\text{T})~{\rm csch}^2{\sqrt{2\over 3\alpha}{\phi-\phi_c\over \text{M}_{\rm pl}}}\biggr], \nonumber
\\
&&r=
{64 \text{T}^2~{\rm csch}^2{\sqrt{2\over 3\alpha}{\phi-\phi_c\over \text{M}_{\rm pl}}} \over 3\alpha }, \nonumber
\\
&&\text{N}_e=
-{3\alpha\over 4\text{T}}\biggl[
\cosh{\sqrt{2\over 3\alpha}{\phi_{\rm end}-\phi_c\over \text{M}_{\rm pl}}}
-\cosh{\sqrt{2\over 3\alpha}{\phi-\phi_c\over \text{M}_{\rm pl}}}\biggr]. \nonumber
\end{eqnarray}
For the E-models, the corresponding results are
\begin{eqnarray}
  n_s &=& 1-\frac{8\text{E}(\text{E}+e^{\sqrt{\frac{2}{3\alpha}}  \frac{\phi-\phi_c}{\text{M}_{\rm pl}}})}{3\alpha(-1+e^{\sqrt{\frac{2}{3\alpha}}  \frac{\phi-\phi_c}{\text{M}_{\rm pl}}})^2} \nonumber ,\\
  r   &=& \frac{64\text{E}^2}{3\alpha(-1+e^{\sqrt{\frac{2}{3\alpha}}  \frac{\phi-\phi_c}{\text{M}_{\rm pl}}})^2}, \nonumber\\
\text{N}_e &=&-\frac{3\alpha}{4\text{E}}\biggl[e^{\sqrt{\frac{2}{3\alpha}}  \frac{\phi_{end}-\phi_c}{\text{M}_{\rm pl}}}-e^{\sqrt{\frac{2}{3\alpha}}  \frac{\phi-\phi_c}{\text{M}_{\rm pl}}}+\sqrt{\frac{2}{3\alpha}}  \frac{\phi-\phi_{end}}{\text{M}_{\rm pl}}\biggr].  \nonumber
\end{eqnarray}
These types of potential in the small $\alpha$ limit are favored by the data from Planck 2015 \cite{Planck:2015xua}.

\begin{figure}[ht!]
 \centering
\includegraphics[width=6cm]{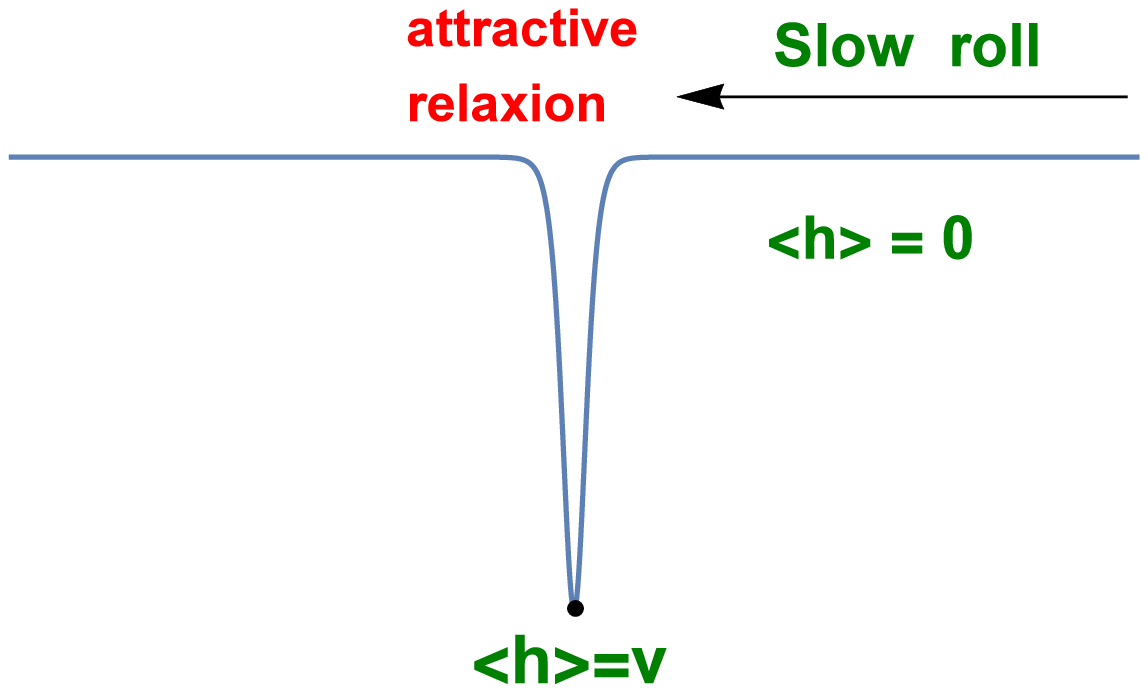}
\includegraphics[width=6cm]{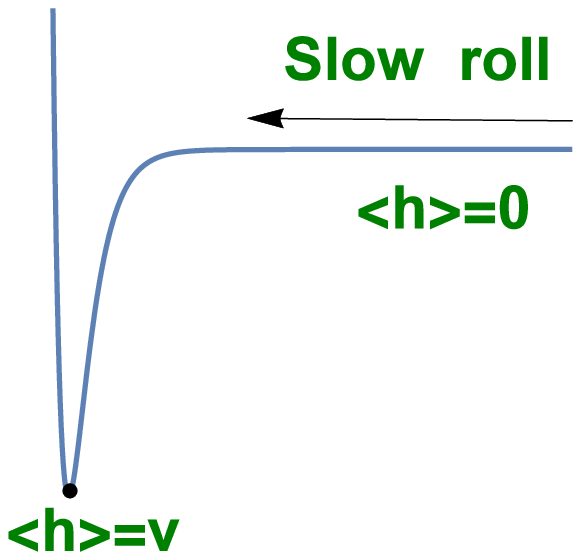}
\caption{(color online) Schematic diagram of the attractive relaxion scenario. The left figure is the case for
 the so-called T-models and the right figure is the case for the E-models.}\label{relaxation}
\end{figure}

Firstly, in the very early universe $\phi \gg  \text{M}/g$, the
effective mass-squared of the Higgs boson $m_h^2$ is positive
and the vacuum expectation value (vev) of the Higgs field is zero.
The final solution is insensitive to
the initial condition of $\phi$ as long as the initial
mass-squared of the Higgs $m_h^2$ is positive,
since it is slow-rolling due to Hubble friction.
The inflaton field $\phi$ drives the slow-roll inflation by the attractive potential $V_{att}=V_{\text{T}}=\text{M}^4\tanh^{2\text{T}}\biggl({\phi-\phi_c\over \sqrt{6\alpha}{\rm M_{\rm pl}}}\biggr)$ or $V_{att}=V_{\text{E}} = \text{M}^{4} \left(1-e^{-{\sqrt {2\over 3 \alpha}}(\phi-\phi_c)/\text{M}_{\rm pl}}\right)^{2\text{E}} $ as shown in Fig.~\ref{relaxation}.
With the evolution of the universe, the inflaton field naturally crosses the
critical point for the Higgs mass where $m_h^2=0$, namely,  a transition point occurs when $\phi=\text{M}/g$.
When the inflaton field across this critical point with $\phi<\text{M}/g$, the mass-squared term $ m_h^2$
transits from a positive value to a negative value, and then the Higgs field acquires a vev.
Thereby, the cosmological inflation can scan the physical mass of the Higgs boson, as
shown in Fig.~\ref{relaxation}.

After reheating and  dissipation\footnote{After the inflation ends,
the reheating process
may be important in these models~\cite{Ueno:2016dim}, but we leave this
for future work.},
the inflaton field $\phi$  is stuck in  the  vicinity of $\phi_c$.
Thus, the Higgs mass evolves from a large field-dependent mass to
the currrent $125$ GeV mass. In the early universe, the field
dependent mass of the Higgs boson is $m_h^2=g^2 \phi^2(t)-\text{M}^2$,
where $\phi(t) \gg  \phi_e$.
During the cosmological evolution, the mass of the Higgs boson
becomes much smaller.
When the inflaton is stuck by the attractive potential of $\phi$ as shown in Fig.~\ref{relaxation},
the Higgs mass is fixed as $\mu^2=g^2 \phi_c^2-\text{M}^2 < 0$. Here, $\mu^2$ is the coefficient of the $h^2$
in the SM with $m_h^2=-2\mu^2=(\rm 125 GeV)^2$.
This provides a cosmological interpretation of the light Higgs mass, and
$\phi_c=\sqrt{\text{M}^2+\mu^2}/{g}$.
For $\rm M=10^6$~GeV, $g=10^{-2}$,
$\phi_c\approx10^8$~GeV.
We call the inflaton field $\phi$  the attractive relaxion, which has the above attractive inflation behavior.

\section{Collider signals at a lepton collider and the muon anomalous magnetic moment}\label{col}
In this section, we discuss the possible collider signals or constraints from particle physics   experiments  for
two simple cases of Higgs portal interactions $f(\phi)h^2/2$.
\subsection{Higgs Invisible Decay}
\begin{figure}[h]
\begin{center}
\includegraphics[width=7cm]{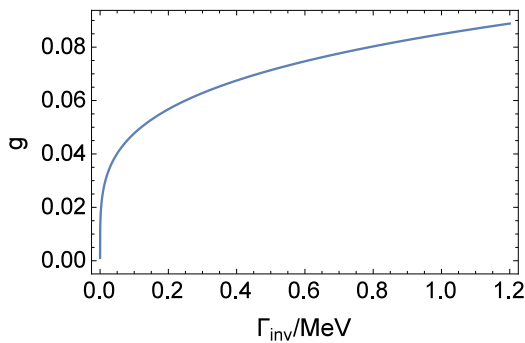}
\caption{  The corresponding Higgs portal coupling to each  Higgs invisible
decay width value.}\label{invisible}
\end{center}
\end{figure}

For the case of $f(\phi)=g^2 \phi^2$ in Eq.~(\ref{effp}), this toy model will contribute to the
Higgs boson invisible decay.
This scenario can be tested at a future lepton collider, such as the circular electro-positron collider (CEPC) by precisely measuring the width of the Higgs invisible decay.
Here, the Higgs invisible decay channel is induced  from
the Higgs portal term $ g^2 h^2 \phi^2/2$
in Eq.(\ref{effp}).
We obtain the following interaction term
\begin{equation}
\mathcal{L}_{h \to \rm \phi \rm \phi}=-g^2 v \phi^2 h,\label{ala}
\end{equation}
which leads to the Higgs invisible decay, and  its
decay width is
\begin{equation}
 \Gamma_{\rm inv}(\rm h \to \phi \phi)=\frac{g^4 v^2}{8 \pi m_h}\sqrt{1-\frac{4m_\phi^2}{m_h^2}}\simeq\frac{g^4 v^2}{8 \pi m_h}.\label{widthinv}
\end{equation}
Figure \ref{invisible} shows the relation between the Higgs portal coupling $g$ and its decay width $\Gamma_{\rm inv}(h)$.
The current Higgs portal coupling from LHC data is constrained as $g<0.088$.

For the future lepton collider,
the expected accuracy for the branching ratio of the Higgs boson invisible decay $\rm BR(h\to inv)$, normalized to $5~\rm ab^{-1}$
is about $0.14\%$ combined~\cite{CEPC-SPPCStudyGroup:2015csa}.
If the signal is not observed at the future CEPC, it will provide  an upper bound for the  Higgs portal coupling of about
$g<0.073$ at the future CEPC~\cite{CEPC-SPPCStudyGroup:2015csa}.

\subsection{Muon anomalous magnetic moment}
For the case of $f(\phi)=g^2 \phi$, the mixing interaction $g^2 v \phi h$ between $\phi$ and $h$ is
induced, and $\vartheta$ is defined as the mixing angle between the inflaton and the Higgs boson.
This  will contribute to the muon anomalous magnetic moment.
Through the effective interaction of the inflaton field with the SM particles by mixing effects, the inflaton $\phi$ can contribute to the muon anomalous magnetic moment.
Up to now, there exists a $3.5 ~\sigma$ deviation between SM predictions and experimental results~\cite{Agashe:2014kda}:
\begin{eqnarray}
 \Delta a_{\mu} = a_\mu^{\rm Exp} - a_{\mu}^{\rm SM} = (236 \pm 87) \times 10^{-11}.
\end{eqnarray}
At the one-loop level, the contribution from the inflaton  $\phi$ to the muon anomalous magnetic moment can be written as
\begin{eqnarray}
 \Delta a_{\mu}^{\rm NP} = \vartheta^2 \frac{\text{G}_{\rm F} m_{\mu}^4}{4\pi^2\sqrt{2}}\int_0^1\frac{y^2(2-y)}{m_\mu^2y^2+m_{\phi}^2(1-y)}dy,
\end{eqnarray}
where $\text{G}_{\rm F}$ is the Fermi constant with $\sqrt{2}\text{G}_{\rm F}=1/v^2$.
Since it needs $ \Delta a_{\mu}^{\rm NP}<\Delta a_{\mu}$, the constraints
on the model parameters can be obtained numerically as
$\vartheta < 0.75$.


\section{A concrete model}\label{sec:model}


In this section, we study a concrete attractive relaxion model, assuming a non-minimal coupling of the inflation field $\phi$ to
gravity, namely $(1/2)\xi R\phi^2$~\cite{Huang:2013oua}.
Such a non-minimal coupling may arise from  some quantum gravity effects, such as the Higgs
field with asymptotically safe gravity~\cite{Cai:2012qi,Cai:2013caa}.
We begin the discussion with the following action~\cite{Huang:2013oua} in the Jordan frame:
\begin{eqnarray}
&&S=
\int d^4 x \sqrt{-\hat{g}}
( \frac{1-\xi \kappa^2 (\phi-\phi_c)^2}{2\kappa^2} R(\hat{g})
\nonumber \\
&&
-\frac{1}{2}(\hat{\nabla} \phi)^2 -\frac{1}{2} (\hat{\nabla}h)^2-V(\phi,h),
\label{lag1}
\end{eqnarray}
where
\begin{equation}\label{effpotential}
V(\phi,h)=\frac{\lambda_{SM}}{4}h^4+\frac{(g^2 \phi^2-\text{M}^2 )}{2}h^2+\frac{\lambda}{4} (\phi-\phi_c)^4,
\end{equation}
and $\kappa^2/8\pi=\text{G}$, which G is Newton's gravitational constant.

Performing the Weyl conformal transformation with
$\Omega^2=1-\xi\kappa^2(\phi-\phi_c)^2$ and
defining a new field $\varphi_1$ to make the kinetic term of $\phi$
be canonical as
\begin{eqnarray}
\varphi_1=\int \sqrt{\frac{1-(1-6\xi)\xi\kappa^2(\phi-\phi_c)^2}{(1-\xi\kappa^2
(\phi-\phi_c)^2)^2}}d\phi,
\label{nonmin2}
\end{eqnarray}
or
\begin{eqnarray}
\frac{d \varphi_1}{d \phi}= \sqrt{\frac{1-(1-6\xi)\xi\kappa^2(\phi-\phi_c)^2}{(1-\xi\kappa^2
(\phi-\phi_c)^2)^2}},
\label{dnonmin2}
\end{eqnarray}
we obtain the following  action in the Einstein frame:
\begin{eqnarray}
&&S=
\int d^4 x \sqrt{-g} ( \frac{1}{2\kappa^2}R- \frac{1}{2}
(\nabla\varphi_1)^2
\nonumber  \\
&& -\frac{1}{2} e^{-2F(\varphi_1)}(\nabla h)^2 -
U(\varphi_1,h) ),
\label{ac}
\end{eqnarray}
with
\begin{eqnarray}
F= \frac12 \ln |1-\xi\kappa^2(\phi-\phi_c)^2|,\nonumber\\
U=e^{-4F(\varphi_1)}V=\frac{V}
{(1-\xi\kappa^2(\phi-\phi_c)^2)^2}.\nonumber
\label{nonmin3}
\end{eqnarray}

We use the longitudinal gauge (Newton gauge),
\begin{equation}\label{long_new}
d s^2= -(1+2 \Phi)d t^2+ a^2(t) (1-2 \Psi) \delta_{ij}d x^i d x^j,
\end{equation}
and  take $\Phi=\Psi$ here.

Firstly, we derive the  background field evolution equations for the
cosmic expansion rate $H =\dot{a}/a$ and homogeneous parts
of scalar fields by variation
of the action in
Eq.(\ref{ac}) :
\begin{eqnarray}
H^2=\frac{\kappa^2}{3} \left(\frac12 \dot{\varphi}_1^2+\frac12 e^{-2F}
\dot{h}^2+U \right),
\label{hubble1}
\end{eqnarray}
\begin{eqnarray}
\dot{H}=-\frac{\kappa^2}{2} \left( \dot{\varphi}_1^2
+e^{-2F}\dot{h}^2 \right),
\label{hubble2}
\end{eqnarray}
\begin{eqnarray}
\ddot{\varphi}_1+3H\dot{\varphi}_1+U_{,\varphi_1}
+F_{,\varphi_1}  e^{-2F} \dot{h}^2=0,
\label{varphi1}
\end{eqnarray}
\begin{eqnarray}
\ddot{h}+3H\dot{h}+
e^{2F}U_{,h}-2F_{,\varphi_1}
\dot{\varphi}_1\dot{h}=0,
\label{varphi2}
\end{eqnarray}
Secondly, following the standard perturbative methods in cosmological inflation~\cite{Mukhanov:1990me}, we
obtain the Fourier-transformed, first-order Einstein equations for the metric and
field fluctuations as
\begin{eqnarray}
&&\ddot{\Phi}+4H\dot{\Phi}+\kappa^2U\Phi  \nonumber \\
&&= \frac{\kappa^2}{2}
\left[ \dot{\varphi}_1\delta\dot{\varphi}_1-
(U_{,\varphi_1}+F_{,\varphi_1}
e^{-2F}\dot{h}^2) \delta\varphi_1+
e^{-2F}\dot{h}\delta\dot{h}-
U_{,h}\delta h \right], \nonumber
\label{perturb4}
\end{eqnarray}
\begin{eqnarray}
&&-\frac{\kappa^2}{2}(\dot{\varphi}_1\delta\dot{\varphi}_1+(3H\dot{\varphi}_1 +U_{,\varphi_1}-F_{,\varphi_1}e^{-2F}\dot{h}^2) \delta\varphi_1 \nonumber \\
&&+e^{-2F}\dot{h}\delta\dot{h}+(U_{,h}+3H\dot{h} e^{-2F}) \delta h ) \nonumber  \\
&&=\left(\frac{k^2}{a^2}-\dot{H}\right)\Phi
,\nonumber
\label{perturb5}
\end{eqnarray}
\begin{eqnarray}
\dot{\Phi}+H\Phi=\frac{\kappa^2}{2}
\left( \dot{\varphi}_1 \delta \varphi_1
+e^{-2F}\dot{h} \delta h \right),\nonumber
\label{perturb1}
\end{eqnarray}
\begin{eqnarray}
&&\delta\ddot{\varphi}_1+ 3H\delta\dot{\varphi}_1
+\left[\frac{k^2}{a^2}+U_{,\varphi_1\varphi_1}-
\left(e^{-2F}\right)_{,\varphi_1\varphi_1}
 \frac{\dot{h}^2}{2} \right] \delta\varphi_1
 \nonumber \\
&=& 4\dot{\varphi}_1 \dot{\Phi}-2U_{,\varphi_1}\Phi- 2F_{,\varphi_1} e^{-2F} \dot{h}
\delta\dot{h}+U_{,\varphi_1 h}\delta h, \nonumber
\label{perturb2}
\end{eqnarray}
\begin{eqnarray}
&&\delta\ddot{h}+(3H-2F_{,\varphi_1} \dot{\varphi}_1)\delta\dot{h}
+\left(\frac{k^2}{a^2}+e^{2F}U_{,hh} \right)
\delta h-2 F_{,\varphi_1} \dot{h} \delta\dot{\varphi}_1 \nonumber \\
&&= 4\dot{h} \dot{\Phi}-2e^{2F}U_{,h}\Phi-e^{2F}
\left( 2F_{,\varphi_1}
U_{,h}+U_{,\varphi_1 h}
-2 F_{,\varphi_1 \varphi_1} \dot{\varphi_1} \dot{h} \right) \delta \varphi_1.\nonumber
\label{perturb3}
\end{eqnarray}
For $\xi >0$, $F=\frac12 \ln (1-\xi\kappa^2(\phi-\phi_c)^2)$,
then $e^{2F}=1-\xi\kappa^2(\phi-\phi_c)^2$,
$e^{-2F}=1/(1-\xi\kappa^2(\phi-\phi_c)^2)$,
$F_{,\varphi_1}=F_{,\phi}/(d \varphi_1/d \phi)$,
and $U_{,\varphi_1}=U_{,\phi}/(d \varphi_1/d \phi)$.
Note that we should be careful in dealing with the
original field $\phi$ and the new field $\varphi_1$.

Write
\begin{equation}
d s^2=a^2(t)[-d\tau^2+(\delta_{ij}+2 B_{ij})d x^i d x^j],
\end{equation}
then we have
\begin{equation}\label{tf}
\ddot{B}+ 3 H \dot{B}+\frac{k^2}{a^2} B=0.
\end{equation}
Setting $\text{M}_{\rm pl}=1$, then the tensor power spectrum is obtained as
\begin{equation}\label{pt}
P_T=\frac{4 k^3}{\pi^2} B^2,
\end{equation}
and the scalar power spectrum is
\begin{equation}\label{ps}
P_S=\frac{ k^3}{2 \pi^2} \zeta^2,
\end{equation}
where the so-called Bardeen parameter is defined by
\begin{equation}\label{zzz}
\zeta=\Phi-\frac{H^2}{\dot{H}}(\Phi+\frac{\dot{\Phi}}{H}).
\end{equation}

\begin{figure}[ht]
 \centering
\includegraphics[width=7cm]{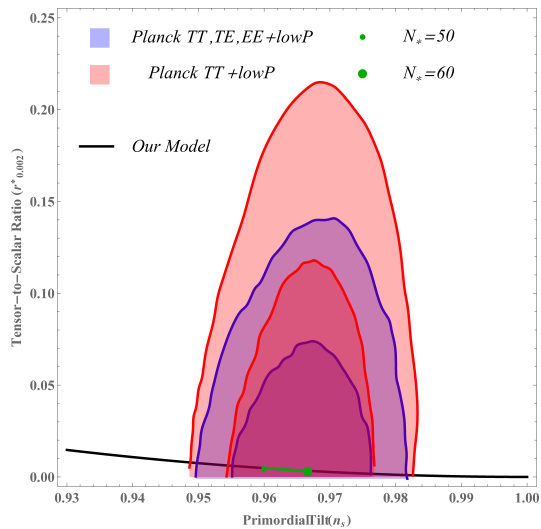}
\caption{(color online)   Marginalized joint $68\%$ and $95\%$ C.L. regions for $n_s$ and $r^*$ from
 Planck 2015~\cite{Ade:2015lrj} compared to the theoretical predictions of  this inflationary model.}\label{nsr}
\end{figure}

Here, we take the single field slow-roll approximation. Then,
the detailed conditions of the cosmological inflation are described by the following slow-roll parameters:
\begin{align}
\label{epsilon}
  \epsilon &
  = \frac{1}{2}\left(\frac{dU/d{\varphi_1}}{U}\right)^2, \\
\label{eta}
  \eta &
  = \frac{d^2U/d{\varphi_1}^2}{U},
\end{align}
where $\epsilon$ and $\eta$ represent the first and second derivatives of the inflation potential in the Einstein frame,
respectively.
The number  of $e$-foldings $\text{N}_e$  is given by
\begin{eqnarray}
 \text{N}_e &=& \frac{1}{\sqrt{2} } \int_{t_i}^{t_f}
 H dt                     \nonumber \\
  \end{eqnarray}
Thus, the  amplitude of density perturbations in $k$-space under the slow-roll approximation is defined by the power spectrum:
\begin{equation}
P_S(k)=A_S \left(\frac{k}{k^*}\right)^{n_s-1},
\end{equation}
where $A_S$ is the scalar amplitude at some ``pivot point" $k^*$, which is given by
\begin{equation}
A_S = \frac{U}{24 \pi^2  \epsilon} \Bigg{|}_{k^*},
\label{Delta}
\end{equation}
which can be measured from   cosmic microwave background radiation(CMB) experiments.
At the leading level, the
scalar spectral index $n_s$ can be written as
\begin{equation}
n_s =  1-6\epsilon+2\eta\,,
\end{equation}
and the corresponding tensor-to-scalar ratio $r^*$  is
\begin{equation}
r^*=  16\epsilon\,.
\end{equation}
Using the recent Planck 2015 data $n_s=0.9655\pm 0.0062 $ and $\ln(10^{10}A_S)=3.089\pm0.036$~\cite{Ade:2015lrj}, we can obtain the constraints on the model parameters,
and fit the combined experimental results of Planck 2015  using this model in the $n_s-r$ plane as shown in Fig. \ref{nsr}, where we
choose the pivot scale $k^*= 0.002 \rm{Mpc^{-1}}$, and $r_{0.002}^* < 0.10$ at
$95\%$C.L. Figure \ref{nsr} shows that our  prediction  is well within the joint $95\%$ C.L. regions.
Roughly speaking, during inflation,  there may exist the effect of entropy perturbation
from the Higgs field.
After reheating, it can be converted to curvature perturbation if the Higgs field couples
to the inflaton field. In particular, if the reheating process is realized by the so-called Higgs
reheating, then this effect would be very manifest. But this only applies to the case when the coupling between the inflation field and the Higgs field is
sufficiently strong. However, the coupling between them is weak in this attractive relaxion model, and thus the result for the curvature perturbation is almost unaffected\footnote{Detailed discussions on the effects of the Higgs field are given in Ref.~\cite{Cai:2013caa}.}.
This non-minimal coupling model just corresponds to a special case of the T-models,
which can explain the light Higgs mass from the cosmological evolution.

\section{Conclusion}\label{sec:sum}

We have put forward  a toy model, which aims at providing a possible interpretation of the Higgs mass
by attractive inflation.
Only the inflaton field is needed and a broad classes of inflation models with
attractive potentials can satisfy the conditions.
This proposal ties the puzzling light mass of the Higgs boson to an attractive inflaton field which plays an important role during
cosmological evolution. The possible collider signals or constraints at the future lepton colliders, the possible constraints
from the muon anomalous magnetic moment and the concrete models were also discussed in detail.
The attractive relaxion idea  here represents
a new interplay between particle physics and
cosmology, and these new ideas of cosmological evolution
would open a new door to  understand some key parameters of particle physics.

\section*{Acknowledgments}
We thank Francis Duplessis for valuable discussions during the early work.
FPH is pleased to recognize the hospitality of
the Department of Astronomy when visiting at University of Science and Technology of China.
FPH and XZ are supported by the NSFC (Grant Nos. 11121092, 11033005, 11375220) and by the CAS pilotB program.
FPH is also supported by the China Postdoctoral Science Foundation under Grant No. 2016M590133.
YFC are supported in part by the Chinese National Youth Thousand Talents Program, by the USTC start-up funding (Grant No. KY2030000049) and by the NSFC (Grant No. 11421303).
HL is supported in part by the NSFC under Grant No. 11033005 and the youth innovation promotion association project and the Outstanding young scientists project of the Chinese Academy of Sciences.

\clearpage
\end{document}